\documentclass[%
 reprint,
 superscriptaddress,
 amsmath,amssymb,
 aps,
 prl,
]{revtex4-2}
\usepackage{siunitx}
\usepackage{xcolor}
\usepackage{lipsum} 

\usepackage{appendix}
\usepackage{braket}
\usepackage{float} 
\usepackage{chemformula}
\usepackage{graphicx}
\usepackage{dcolumn}
\usepackage{bm}
\usepackage{setspace}
\usepackage{mathtools}
\usepackage{braket}
\usepackage{graphicx}
\usepackage{dcolumn}
\usepackage{bm}
\usepackage[colorlinks=true, citecolor=blue, linkcolor=blue, urlcolor=blue]{hyperref}
\usepackage{multirow}

\makeatletter
\newcommand*{\addFileDependency}[1]{
  \typeout{(#1)}
  \@addtofilelist{#1}
  \IfFileExists{#1}{}{\typeout{No file #1.}}
}
\makeatother

\setcitestyle{super}

\begin{document}

\title{Response regimes in on-chip THz spectroscopy
}

\newcommand{\affiliationMPSD}{
Max Planck Institute for the Structure and Dynamics of Matter,
Luruper Chaussee 149, 22761 Hamburg, Germany
}

\newcommand{\affiliationCFEL}{
Center for Free-Electron Laser Science (CFEL),
Luruper Chaussee 149, 22761 Hamburg, Germany
}

\newcommand{\affiliationETH}{
Center for Free-Electron Laser Science (CFEL),
Luruper Chaussee 149, 22761 Hamburg, Germany
}

\newcommand{\CU}{Department of Physics, Columbia University,
538 West 120th Street, New York, NY 10027, USA}

\newcommand{\PKS}{Max Planck Institute for the Physics of Complex Systems, Nöthnitzer Straße 38, 01187 Dresden,
Germany}

\author{Gunda Kipp}
\thanks{Corresponding author: \href{gukipp@phys.ethz.ch}{gukipp@phys.ethz.ch}}
\affiliation{\affiliationMPSD} 
\affiliation{\affiliationCFEL}

\author{Marios H. Michael}
\affiliation{\PKS}
\affiliation{\affiliationMPSD}
\affiliation{\affiliationCFEL}

\author{Alexander M. Potts}
\affiliation{\affiliationMPSD}
\affiliation{\affiliationCFEL}
\affiliation{\CU}

\author{Dorothee Herrmann}
\affiliation{\affiliationMPSD}
\affiliation{\affiliationCFEL}

\author{Toru Matsuyama}
\affiliation{\affiliationMPSD}
\affiliation{\affiliationCFEL}

\author{Guido Meier}
\affiliation{\affiliationMPSD}
\affiliation{\affiliationCFEL}

\author{Matthew W. Day}
\affiliation{\affiliationMPSD}
\affiliation{\affiliationCFEL}
\affiliation{\CU}

\author{Hope M. Bretscher}
\affiliation{\affiliationMPSD}
\affiliation{\affiliationCFEL}
\affiliation{\CU}

\author{James W. McIver}
\thanks{Corresponding author: \href{jm5382@columbia.edu}{jm5382@columbia.edu}}
\affiliation{\affiliationMPSD}
\affiliation{\affiliationCFEL}
\affiliation{\CU}
\date{\today}

\begin{abstract}
On-chip THz spectroscopy enables quantitative measurements of the optical conductivity of sub-wavelength 2D materials by tightly confining THz fields in metallic transmission line structures interfaced to the material. However, because the probed structures are smaller than the THz wavelength, finite-size and environmental effects can strongly influence the measured response. Here, we identify the conditions under which a metallic sample exhibits a genuine Drude response and when finite-size and environmental effects must be considered. We further introduce and characterize an additional regime, the Phantom-Drude response, which mimics Drude behavior but instead originates from the superposition of multiple finite-momentum plasmonic resonances. If unrecognized, this regime can lead to misinterpretation of intrinsic material properties. We systematically show how the Phantom-Drude response can emerge and demonstrate its sensitivity to sample dimensions, transmission line geometry, material shape, and gate properties, providing practical guidelines to avoid this regime in future on-chip THz measurements.

\end{abstract}
\maketitle

\paragraph{\textbf{Introduction.}}

Two-dimensional (2D) materials and van der Waals (vdW) heterostructures host a variety of emergent, gate-tunable phenomena \cite{cao2018unconventional,mak2022semiconductor,corrininsulinTBG2020,cao2020strange,jindal2023coupled,xia2025superconductivity,basov2020polariton}. Many of these, such as superconductivity, correlated insulating states, and strange-metal phases, arise at \SI{}{\micro eV} - meV energy scales \cite{xie2019spectroscopic,armitage2009electrodynamics,zibrov2017tunable,Nuckolls2024,park2023observation,Balents2020}, corresponding to terahertz (THz) frequencies. While conventional far-field THz techniques can access these frequencies, they are diffraction-limited and cannot resolve micron-scale samples smaller than the THz wavelength. This limitation has spurred the development of near-field THz techniques capable of probing sub-wavelength structures \cite{hillenbrand2025visible,chen-ssnom2012,ultrafastSTMcocker2013,vonhoegen2025visualizingterahertzsuperfluidplasmon,spintronicTHzemitter2016,Terhertzemissionfromgiant2025,helmrich2025cavitydrivenattractiveinteractionsquantum}. Crucially, signals from sub-wavelength samples can differ qualitatively from far-field measurements of large-area samples, as finite-size and environmental effects dominate at small scales \cite{Hillenbrand2025}.

\begin{figure}[t!]
    \centering
    \includegraphics[width = 1. \linewidth]{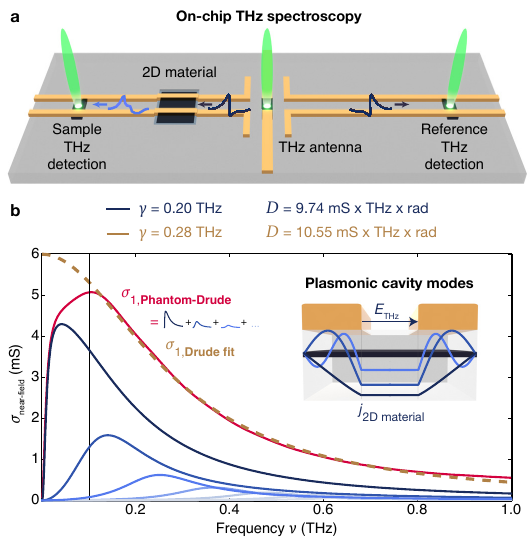}  
    \caption{\textbf{Probing the near-field optical conductivity of 2D materials with on-chip THz spectroscopy: a} Schematic of the on-chip THz circuit architecture: Two identical THz pulses (dark blue) propagate along symmetric transmission lines; one interacts with a 2D material (blue gradient) and is detected on the left, while the other serves as a reference on the right. Comparing both signals yields the material’s near-field optical conductivity. \textbf{b} 2D materials embedded in on-chip THz spectroscopy circuitry form plasmonic self-cavities, featuring one or more resonances in the near-field optical conductivity spectrum \cite{kipp2024cavity}. These resonances arise from finite-momentum standing waves of current density spanning the sample, as illustrated for the three lowest frequency modes of $\sigma_{\text{near-field}}$ (blue). When the resonance linewidth $\gamma$ and the frequency resolution are broader than the mode spacing, overlapping resonances merge into a broad spectrum that mimics a Drude response (red), despite being fundamentally different. A Drude fit to the red spectrum (gold dashed curve) yields misleading parameters: the extracted scattering rate $\gamma$ and Drude weight $D$ (gold) differ significantly from the intrinsic material values (blue). We discuss the origin of this effect and mitigation strategies in this work. Sample parameters for calculations: $W_0=$~\SI{0}{\micro\meter} (2D material width exceeding the metal strips), $W_1=$~\SI{10}{\micro\meter} (width of each metal trace) and $W_2=$~\SI{10}{\micro\meter} (width of the gap between strips), $d_{\text{gold}}=$~\SI{285}{nm} (gold thickness), $d_{\text{hBN}}=$~\SI{20}{nm} (hBN thickness), $d_{\text{sapphire}}=$~\SI{2}{mm} (sapphire substrate thickness), $D=\SI{9.74}{mS\times THz \times rad}$ (Drude weight of the 2D material) and $\gamma=\SI{0.2}{THz}$ (scattering rate of the 2D material).}
    \label{fig:fig1}
\end{figure}

Of the available near-field methods, on-chip THz spectroscopy \cite{kipp2024cavity,zhao2023observation,gallagher2019quantum,seo2024onchipterahertzspectroscopydualgated,chen2024directmeasurementterahertzconductivity,potts2023chip,island2020chip,sterbentz2023chip} is uniquely suited to probe the influence of sub-wavelength THz effects on a sample’s response, because there is an analytical mapping between the near-field signal and the intrinsic conductivity \cite{marioskipp2025}. In this approach, vdW heterostructures are integrated into ultrafast optoelectronic circuits \cite{yoshioka2025,Joe25,li2025purcellenhancementphotogalvaniccurrents,Li:25,Mosley:24,yoshioka2024chip,wang2023superconducting,mciver2020light,Jhuria2020,karnetzky2018towards,YangYang2017,zhong2008terahertz} that confine THz fields using lithographically defined metallic transmission lines (Fig.~\ref{fig:fig1}a). Femtosecond laser pulses excite photoconductive switches that launch and detect THz signals, enabling time-resolved transmission or reflection measurements. By using an \textit{in situ} calibration arm \cite{Katy2025} and incorporating circuit geometry into the analysis \cite{kipp2024cavity}, the real and imaginary parts of the complex near-field conductivity, $\sigma_{\text{near-field}}$, can be extracted, which directly reflects both the intrinsic electronic response and local dielectric environment.

Previous on-chip THz spectroscopy works have interpreted the optical conductivity of metallic samples in one of two ways. One group of studies attributed the spectral features to a Drude response \cite{gallagher2019quantum,chen2024directmeasurementterahertzconductivity,seo2024onchipterahertzspectroscopydualgated,sudichenwigneronchip2025}, where zero-momentum charge transport is governed by electron scattering and dissipation. Another set of studies reported finite-momentum plasmonic resonances \cite{kipp2024cavity,zhao2023observation,pierce2025imagingpropagatingterahertzcollective,Potts2026}, arising from standing wave patterns of current density that are reflected off of both the edges of the transmission lines and the edges of the THz sub-wavelength 2D material. Here, the material effectively forms a plasmonic \textit{self-cavity} for THz light \cite{marioskipp2025} (see inset of Fig.~\ref{fig:fig1}b). The coexistence of these two interpretations in the literature highlights the importance of identifying the conditions under which each applies to ensure the accurate extraction of material parameters.

In this work, we analytically examine the factors that give rise to Drude or cavity responses and identify an intermediate regime, the \textit{Phantom-Drude} response, which, if unrecognized, can lead to erroneous interpretations. While it is known that time-domain reflections or poor choice of window functions can obscure spectral features from collective modes \cite{marulanda2025windowingterahertztimedomainspectroscopy}, we show that the Phantom-Drude response arises from a fundamentally different mechanism: the superposition of multiple, inhomogeneously broadened finite-momentum plasmonic collective modes. As illustrated in Fig.~\ref{fig:fig1}b, the overlap of several distinct plasmon resonances (blue) can produce a broad, featureless spectrum (red). Such a spectrum appears Drude-like when the measurement bandwidth is limited to be in the range of \SIrange{0.1}{1}{THz}, as is often the case experimentally. However, fitting such a spectrum with a Drude model (gold dashed line), leads to the extraction of material parameters that can be \SIrange{10}{400}{\%} off from their intrinsic values.

We analyze three mechanisms underlying the Phantom-Drude regime, arising from the transmission line design, sample geometry, and choice of gate materials, and provide practical design guidelines to mitigate them. These insights enable future on-chip THz spectroscopy studies to reliably probe the intrinsic properties of vdW heterostructures.\\

\begin{figure*}[t!]
    \centering
    \includegraphics[width=1. \linewidth]{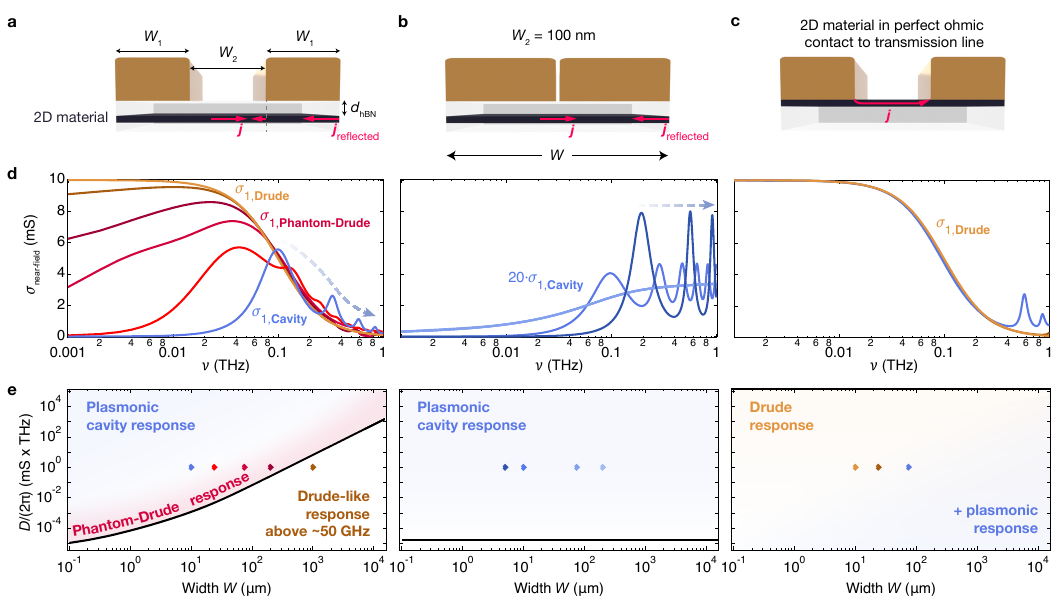}    \caption{\textbf{Impact of transmission line design and material conductivity on the measured response:} Three transmission line designs are investigated. \textbf{a} In the first design, a 2D material is electrically insulated from the transmission line by a hBN layer of thickness $d_{\text{hBN}}=$~\SI{20}{nm}. The structure is embedded within a transmission line of total width $W=2W_1+W_2$ with $W_1=W_2$. No material is extending beyond the transmission line ($W_0=0$, see SI). \textbf{b} In the second design, the transmission line has a small gap of width $W_2=$~\SI{100}{nm} that is kept constant while tuning the overall width $W$. \textbf{c} In the third design, the hBN spacer is removed and the 2D material is placed in perfect ohmic contact to the transmission line of total width $W=2W_1+W_2$ with $W_1=W_2$. \textbf{d} displays the real parts of the near-field optical conductivity as function of frequency corresponding to the markers in panels (e). \textbf{e} The nature of the response (whether plasmonic cavity, Phantom-Drude, or Drude) is calculated as a function of $W$ and the intrinsic Drude weight $D$ of the material, assuming a fixed scattering rate of \SI{0.1}{THz}, for each of the three designs illustrated in (a)–(c). For the design shown in (a), any of the three responses (plasmonic cavity, phantom-Drude, or Drude) can be observed depending on $W$ and $D$. In contrast, the design in (b) permits only cavity responses or broad spectral features that do not resemble a Drude response. The design in (c), featuring an ohmic contact, consistently produces a Drude response regardless of $W$ or $D$. Additionally, plasmonic resonances can appear on top of the Drude response for large or low-conductive samples due to the formation of unscreened plasmons in the transmission line gap (see SI). The black curves in the left (data adapted from Fig.~2d of Ref.~\cite{kipp2024cavity}) and center panel mark parameter sets ($D$, $W$) for which the real part of the near-field optical conductivity, measured via on-chip THz spectroscopy, deviates less than \SI{10}{\%} from the intrinsic 2D conductivity at \SI{10}{GHz}. Notably, the phantom-Drude regime (red-shaded area in (e)) expands with increasing linewidth, while the black curve remains fixed. Note that for design (c), full 3D electromagnetic simulations were used in place of the analytical theory (see SI).}
    \label{fig:fig2}
\end{figure*}

\paragraph{\textbf{Mechanism 1: Reflection of current set by transmission line design.}}

The analytical framework for on-chip THz spectroscopy \cite{kipp2024cavity,marioskipp2025} calculates the effective near-field conductivity for structures where an insulating layer separates the 2D material from the transmission line, capturing how the response is shaped by both the sample geometry and the local dielectric environment. The theory shows that the THz field, largely confined between the transmission line metal strips, drives in-plane currents in the 2D metallic layers perpendicular to the pulse propagation direction (see inset of Fig.~\ref{fig:fig1}b). At screened regions beneath the metal traces ($W_1$), the unscreened gap between the traces ($W_2$), and the material regions extending beyond the width of the transmission line ($W_0$), these currents undergo partial reflection and transmission (see Fig.~\ref{fig:fig2}a), due to the different dispersion relations of screened and unscreened plasmons.

In the screened regions, the currents form plasmons with low-frequency resonances $\omega_1=\sqrt{Dq_1^2d/\epsilon_0\epsilon_1}$, while in the unscreened regions the resonances shift to higher frequencies $\omega_{0,2}=\sqrt{Dq_{0,2}/\epsilon_0(\epsilon_1+\epsilon_2)}$, and $\omega=2\pi\nu$. Here, $D$ is the Drude weight, defining 
$\sigma_{\text{Drude}} = D/\left(2\pi(\gamma- i\nu)\right)$ with scattering rate $\gamma$, $\epsilon_0$ is the vacuum permittivity, $d$ and $\epsilon_1$ denote the thickness and dielectric constant of the insulating layer, and $\epsilon_2$ the substrate dielectric. The plasmon momenta are set by the widths of the corresponding regions ($q_{0,1} \approx a\pi/(2W_{0,1}$) and $q_2\approx a\pi/W_2$, with $a=1,3,5,\dots$). Each permitted momentum, and therefore each excited plasmon wavelength (see inset of Fig.~\ref{fig:fig1}b), selects a distinct point on the dispersion relation, giving rise to a corresponding resonance. 

Accounting for all boundary conditions (Refs.~\cite{kipp2024cavity,marioskipp2025}), the overall effective near-field conductivity of the entire heterostructure is given by:

\begin{equation}
    \sigma_{\text{near-field}}=\sigma_{\text{Drude}}\left( 1+F\right),
    \label{generalformula}
\end{equation}

\noindent
with the feedback factor for $W_0=0$ (see SI for $W_0 \neq 0$)\cite{marioskipp2025}

\begin{equation}
    F=\frac{q_{\text{2}}\text{ sinc}\left(q_{2}W_2/2 \right)}{-q_{\text{2}}\cos(q_{\text{2}}W_2/2)+q_{\text{1}}\sin(q_{\text{2}}W_2/2)\tan(q_{\text{1}}W_1)}.
    \label{feedbackfactor}
\end{equation}

\noindent
Here, the finite momenta in the dissipative limit are given by $q_1=\sqrt{\epsilon_0\epsilon_1\omega(\omega+2\pi i\gamma)/(dD)}$ and $q_2=\epsilon_0(\epsilon_1+\epsilon_2)\omega(\omega+2\pi i\gamma)/D$. As shown in Equations~\ref{generalformula} and \ref{feedbackfactor}, the feedback factor $F$ alone determines whether the effective near-field conductivity follows a Drude response ($F \approx 0$) or exhibits self-cavity effects with well-defined plasmonic resonances ($F \neq 0$). Since $F$ depends on the 2D material’s Drude weight $D$, the sample geometry ($W_1$, $W_2$, $d$, and $W_0$), the surrounding dielectric environment ($\epsilon_1$ and $\epsilon_2$), and the scattering rate $\gamma$, the type of response observed in on-chip THz spectroscopy is entirely determined by the sample’s conductivity, dimensions, and transmission line design.

We therefore theoretically examine three distinct transmission line designs to highlight how they govern the near-field response (see Fig.~\ref{fig:fig2}): (a) a transmission line with equal metal trace and gap widths ($W_1 = W_2$), where an insulating hBN layer electrically isolates the 2D material from the transmission line, (b) a design with a small gap width ($W_2 \ll W_1$), also incorporating an hBN spacer, and (c) a configuration identical to (a) but without an hBN layer, allowing the 2D material to form an ohmic contact with the transmission line. We then calculated the near-field optical conductivity for a metallic sample with Drude weight $D = \SI{6.28}{mS\times THz\times rad}$ and scattering rate $\gamma = \SI{0.1}{THz}$. These values are representative of a thin graphite or gated graphene flake, providing an illustrative example of materials that display a Drude response in the far-field but exhibit plasmonic behavior in the near-field \cite{horng2011drude,kipp2024cavity}. We evaluated the response over a range of sample widths $W$ (see Fig.~\ref{fig:fig2}d) and mapped the resulting response types as a function of both $D$ and the total sample width $W$ for each transmission line geometry (see Fig.~\ref{fig:fig2}e). The black contours in Fig.~\ref{fig:fig2}e indicate the boundary within which the near-field optical conductivity deviates by less than \SI{10}{\%} from the intrinsic Drude conductivity at \SI{10}{GHz}, corresponding to $|F| < 0.1$ at \SI{10}{GHz} and $F \approx 0$ at frequencies within the typical experimental bandwidth (see Fig.~S18 of Ref.~\cite{kipp2024cavity}). We chose this benchmark because a Drude response is recovered above \SI{10}{GHz} and scattering rates and Drude weights can be accurately extracted.

For design (a), plasmonic resonances appear prominently in the near-field optical conductivity of a sample with Drude weight $D = \SI{6.28}{mS\times THz \times rad}$ and a narrow total width $W = 2W_1 + W_2 = \SI{10}{\micro\meter}$ (blue trace, Fig.~\ref{fig:fig2}d). As $W$ increases, the resonances shift to lower frequencies (red traces), and their superposition produces a Phantom-Drude response that persists up to $W = \SI{200}{\micro\meter}$ for this Drude weight. Only for widths approaching $W \sim \SI{1}{mm}$ (dark golden trace) does the spectrum begin to resemble a Drude response above $\sim$~\SI{50}{GHz}. Because current on-chip THz circuits typically lack reliable sensitivity below approximately \SIrange{50}{100}{GHz} due to time-domain reflections, such a spectrum could be fit with a Drude model to reliably extract intrinsic material parameters. 

However, widths approaching \SI{1}{mm} remain challenging to realize for vdW heterostructures. A Drude-like response at experimentally accessible dimensions can instead be achieved only when the Drude weight is sufficiently low, as both increasing $W$ and decreasing $D$ shift the plasmonic resonances to lower frequencies (Fig.~\ref{fig:fig2}e, left). Even in this case, however, care must be taken to confirm that the observed spectrum reflects a true Drude response rather than a Phantom-Drude one (see SI).

One key aspect for the appearance of the Phantom-Drude response in design (a) is the momentum-dependent amplitude decay of the plasmonic self-cavity modes: lower-frequency modes exhibit higher amplitudes, while higher-frequency modes decay in amplitude, as illustrated with the blue dashed arrow in the left panel of Fig.~\ref{fig:fig2}d. As the self-cavity modes shift to lower frequencies with increasing sample width or decreasing Drude weight, their superposition can mimic a Drude spectrum, where conductivity decreases with frequency.

In contrast, the 2D material in design (b) is predominantly screened due to the narrow transmission line gap ($W_2\ll W_1$). As a result, the self-cavity modes exhibit the characteristics of screened 2D plasmons with uniform amplitudes across all excited momenta (see dashed blue arrow in center panel of Fig.~\ref{fig:fig2}d). Consequently, the resulting superposition of plasmonic modes does not mimic a Drude response, even as the sample width $W$ is increased or the Drude weight $D$ is decreased (see Fig.~\ref{fig:fig2}d,e, center panels). We note that the model does not account for Landau damping of high-momentum modes or other momentum-dependent scattering mechanisms, which could suppress or broaden higher-momentum excitations.

While we focus here on two representative cases, design (a) with $W_1=W_2$ and $W_0=0$, and design (b) where $W_2\ll W_1$ and $W_0=0$, we have also simulated a wide range of geometries with $W_2 \neq W_1$ and $W_0\neq 0$ (see SI). These simulations reveal that the Phantom-Drude response persists in all cases, disappearing only in the limit $W_2 \ll W_1$ and $W_0=0$, corresponding to design~(b).

Finally, when a 2D material is placed in direct ohmic contact with the transmission line, as shown in design (c) of Fig.~\ref{fig:fig2}c, a true Drude response ($F=0$ at all frequencies) is observed regardless of the sample width or Drude weight (see Fig.~\ref{fig:fig2}d,e, right panels). In the 2D material, the THz-driven current flows primarily in and out of the metal traces to which the sample is ohmically contacted, producing a Drude response. Notably, plasmons can also form in the unscreened $W_2$ region, adding resonant features on top of this Drude background. However, such plasmonic resonances fall within the experimental bandwidth only for sufficiently large samples or for materials with low conductivity (see SI).

Based on these three design characterizations, the potential for a Phantom-Drude response arises predominantly in a transmission line geometry where $W_2 \not\ll W_1$ and $W_0 \not\ll W_1$, so that the plasmons that form in the unscreened regions generate self-cavity resonances with momentum-dependent amplitudes. This also requires that the 2D material is electrically isolated from the transmission line by an insulating layer such as hBN (design (a)). While design (b) can also produce a broad spectrum, this design has a clear cavity response and is therefore less prone to misinterpretation. Design (c), where the 2D material is in direct ohmic contact with the transmission line, offers the most reliable conditions for observing a true Drude response, assuming a perfect ohmic contact can be made to the sample. However, for certain sample widths or conductivities, plasmonic resonances can emerge on top of the Drude response from the $W_2$ region. Device dimensions should therefore be chosen to prevent the plasmons from obscuring or inhomogeneously broadening the Drude response (see SI).

Having established how sample size and Drude weight govern the appearance of a Phantom-Drude response in design (a), we now examine the role of linewidth. In principle, self-cavity modes with infinitely narrow linewidths would produce clearly distinguishable resonances, eliminating confusion with a Drude response (see SI). However, as the linewidth increases, whether due to enhanced intrinsic electron scattering or inhomogeneous broadening, the ability to resolve individual modes diminishes, making a broad spectrum increasingly difficult to interpret. In the following, we examine two key mechanisms responsible for inhomogeneous linewidth broadening in on-chip THz spectroscopy.\\

\paragraph{\textbf{Mechanism 2: Inhomogeneous linewidth broadening through irregularly shaped 2D materials.}}

\begin{figure}[t!]
    \centering
    \includegraphics[width = 1. \linewidth]{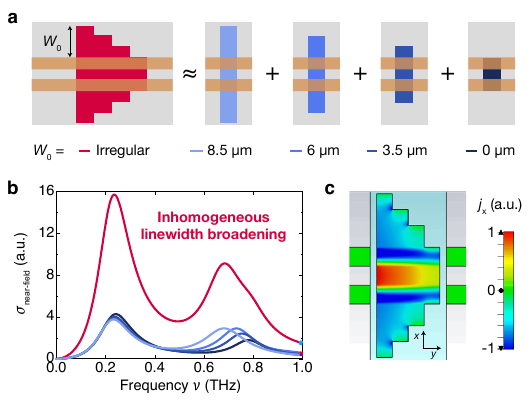}    \caption{\textbf{Inhomogeneous linewidth broadening due to irregularly shaped 2D materials: a} An irregularly shaped 2D material embedded in on-chip THz spectroscopy can be decomposed into slices of fixed width (here, \SI{2.5}{\micro\meter}). Since the frequencies of the self-cavity modes strongly depend on the local width $W_0$ of the part of the 2D material extending the transmission line, the overall near-field optical conductivity spectrum of the irregular flake can be approximated as the sum of the responses from each individual slice. \textbf{b} Full 3D electromagnetic simulations are performed for both the entire irregular flake and its constituent slices. \textbf{c} Full 3D electromagnetic simulations of the current density distribution in the irregularly shaped sample at \SI{0.686}{THz} and a phase of \SI{103}{^\circ}. Distinct plasmonic modes of different wavelengths, and thus different resonance frequencies, form in different slices of the sample, whose combined response gives rise to the inhomogeneously broadened high-frequency peak shown in (b). Sample parameters: $W_1=W_2=$~\SI{3}{\micro\meter}, $D=$~\SI{24.5}{mS\times THz\times rad}, $d_{\text{hBN}}=$~\SI{25}{nm} and $\gamma =1.5/(2\pi)$~THz. }
    \label{fig:fig3}
\end{figure}


Since plasmonic oscillations of the current density extend from one sample edge to the other, the resonance frequencies of the self-cavity modes are determined not only by the transmission line geometry but also by the overall sample width (the length is largely irrelevant; see Ref.~\cite{kipp2024cavity}). If the sample extends irregularly beyond the transmission line, with non-rectangular boundary conditions, different regions along the THz pulse propagation direction will support self-cavity modes with varying wavelengths and resonance frequencies. As illustrated in Fig.~\ref{fig:fig3}, such a non-uniform 2D material can be conceptually divided into slices, each exhibiting a distinct plasmonic wavelength and thus resonance frequency. The superposition of these detuned modes leads to inhomogeneous broadening, and, if the overall resonance frequencies are sufficiently low, this can give rise to a Phantom-Drude response.

This issue can be mitigated by using precisely shaped, rectangular 2D materials, achievable through anodic oxidation lithography \cite{AFMlitho2018} or etching methods.\\

\paragraph{\textbf{Mechanism 3: Inhomogeneous linewidth broadening through hybridization with gate plasmons.}}

To fully explore the rich physics of vdW materials, electrostatic gates are essential for enabling \textit{in situ} tuning of carrier density. However, incorporating such gates into on-chip THz spectroscopy introduces additional complexity. Since electrostatic gating requires the gate material to be at least weakly conductive to facilitate the formation of mirror charges and modulate the carrier density, it unavoidably introduces THz absorption. To mitigate this effect, different groups have adopted a variety of strategies. Some have employed low-conductive gates, which exhibit a weak response in the THz regime \cite{gallagher2019quantum,chen2024directmeasurementterahertzconductivity,hesp2022wse2, sterbentz2024gating}. Others have implemented photogating techniques, in which a semiconducting transition-metal dichalcogenide (TMD) gate becomes conductive only under optical excitation \cite{seo2024onchipterahertzspectroscopydualgated}. Alternatively, metallic graphite gates have been integrated into on-chip THz circuitry with a dedicated reference arm (Fig.~\ref{fig:fig1}a) \cite{kipp2024cavity,Katy2025}, allowing direct characterization of the metallic gate response when the vdW material is at charge neutrality or in an insulating state.

\begin{figure*}[t!]
    \centering
    \includegraphics[width = 1. \linewidth]{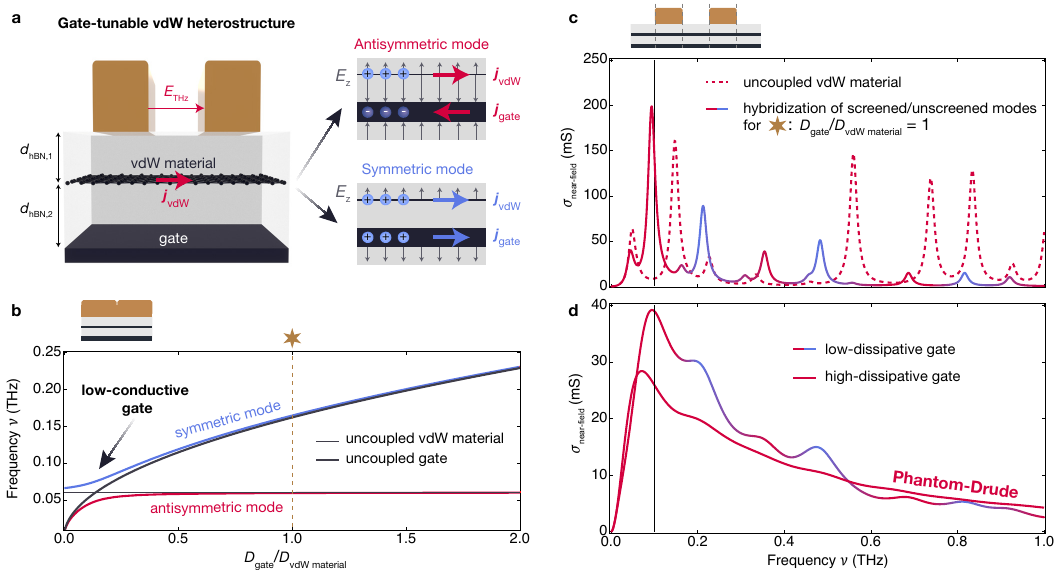}    \caption{\textbf{The challenge of low-conductive gates: a} In on-chip THz spectroscopy of gate-tunable vdW heterostructures, plasmonic modes in the vdW layer and nearby gate strongly couple due to their nanometer separation, forming hybrid symmetric (in-phase) and antisymmetric (out-of-phase) excitations. \textbf{b} Calculated resonance frequencies of hybrid and uncoupled modes versus gate-to-vdW material Drude weight ratio. Highly conductive gates yield weak hybridization, while low-conductive gates cause strong mode mixing and large frequency shifts. \textbf{c} Near-field conductivity spectra for gate-tunable vdW heterostructures in a transmission line of design (a) (see Fig.~\ref{fig:fig2}a). Two cases are compared: a perfectly screened vdW layer with a thick metallic gate (dashed red) and a vdW layer with a lower conductive gate where $D_{\text{gate}}/D_{\text{vdW material}} = 1$ (red–blue). In the latter, antisymmetric (vdW material-like, red) and symmetric (gate-like, blue) hybrid plasmon modes appear. The antisymmetric modes shift to lower frequencies due to coupling with the symmetric modes, a result of resonant interaction at the dielectric boundaries provided by the metal trace edges, as described in Ref.~\cite{kipp2024cavity}. \textbf{d} When the scattering rate is increased to $\gamma_{\text{gate}} = \gamma_{\text{vdW material}}$, distinct spectral features from multiple resonances remain visible (red-blue curve). However, if the gate's scattering rate becomes much larger than that of the vdW material, both the symmetric (blue) and antisymmetric (red) modes broaden significantly, merging into a smooth, featureless Phantom-Drude response (red curve). Sample parameters for panel (b): momentum $q=\pi/(\SI{39}{\micro\meter})$, $D_{\text{vdW material}}=$~\SI{43.35}{mS\times THz \times rad}, $d_{\text{hBN,1}}=$~\SI{20}{nm} and $d_{\text{hBN,2}}=$~\SI{100}{nm}. Sample parameters for panels (c) and (d): $W_0=W_1=W_2=$~\SI{13}{\micro\meter}, $d_{\text{hBN,1}}=$~\SI{20}{nm}, $d_{\text{hBN,2}}=$~\SI{100}{nm}, $D_{\text{vdW material}}=D_{\text{gate}}=$~\SI{34.05}{mS\times THz \times rad}. Scattering rates in (c): $\gamma_{\text{gate}}=\gamma_{\text{vdW material}}=\SI{0.02}{THz}$. Scattering rates in (d): $\gamma_{\text{gate}}=\gamma_{\text{vdW material}}=\SI{0.12}{THz}$ (red-blue) and $\gamma_{\text{gate}}=5\times\gamma_{\text{vdW material}}=\SI{0.60}{THz}$ (red).}
    \label{fig:fig4}
\end{figure*}

A common strategy to mitigate gate contributions to the THz signal is to subtract them. However, as we demonstrate below, this approach can introduce complications. Plasmons can form in both vdW material and gate that hybridize strongly to form symmetric (in-phase) and antisymmetric (180 degrees out-of-phase) plasmonic modes (see Fig.~\ref{fig:fig4}a), that, depending on the conductivities and separations of the two layers, can significantly differ in resonance frequencies and spectral weight from their uncoupled counterparts (see Fig.~\ref{fig:fig4}b). In particular, when the gate has a low Drude weight ($D_{\text{gate}}/D_{\text{vdW material}} \ll 1$), an avoided crossing appears where the uncoupled gate and vdW material modes would otherwise cross in frequency. This avoided crossing signifies strong hybridization, highlighting that low-conductive gates can complicate the extraction of intrinsic vdW material properties. For example, a low-conductive gate's plasmonic resonance cannot be removed from a measured conductivity spectrum by simple subtraction, as such a procedure can even produce unphysical negative effective conductivities (see SI).

In contrast, when the layers are well separated by a thick hBN layer and/or the gate is much more conductive than the 2D material, the modes behave approximately as independent gate and 2D material plasmons (see Fig.~\ref{fig:fig4}b and also Ref.~\cite{kipp2024cavity}).

However, symmetric and antisymmetric modes only behave as uncoupled gate and 2D material plasmons if the gate-tunable heterostructure is placed in a homogeneously screened environment, such as in transmission line design (b) (see Fig.~\ref{fig:fig2}b). When the system follows transmission line design (a) (Fig.~\ref{fig:fig2}a), the situation becomes much more complex. Symmetric and antisymmetric modes can form both in the screened regions beneath the metal traces and also in the unscreened regions in the transmission line gap and outside the transmission line, with different dispersion relations (see SI). As the antisymmetric unscreened modes in the $W_0$ and $W_2$ regions (see Fig.~\ref{fig:fig4}a) have comparable resonance frequencies to the screened antisymmetric and symmetric modes in the $W_1$ regions, they can resonantly enhance the coupling of the screened modes, even of different momenta \cite{kipp2024cavity}. In some cases, this can push the system into the \textit{ultrastrong coupling} regime \cite{forn2019ultrastrong,frisk2019ultrastrong}, where the hybrid mode splitting exceeds \SI{10}{\%} of the original resonance frequencies. 

This type of coupling has a key consequence: in a homogeneously screened heterostructure, symmetric and antisymmetric plasmonic modes can closely resemble the uncoupled modes of the gate and the vdW material. As shown in Fig.~\ref{fig:fig4}b, when the gate and vdW material have comparable Drude weights ($D_{\text{gate}}/D_{\text{vdW material}} \approx 1$), the resonance frequency of the antisymmetric (vdW material-like) mode remains essentially unchanged, even as the gate conductivity increases beyond this ratio. In contrast, this behavior changes dramatically in a micropatterned dielectric environment, such as that introduced by transmission line design (a). As illustrated in Fig.~\ref{fig:fig4}c, the antisymmetric mode, originally resembling that of the uncoupled vdW material (dashed red), is shifted to lower frequencies (solid red) due to hybridization with symmetric, gate-like modes (solid blue), even at $D_{\text{gate}}/D_{\text{vdW material}} = 1$.

Furthermore, hybridization with the gate affects not only the resonance frequencies but also the linewidths of the vdW material–like modes. When the gate exhibits high scattering rates, the increased damping broadens not only the symmetric gate-like mode but also the antisymmetric mode associated with the 2D material, leading to an artificially broadened spectral response of the vdW material. When many such broadened hybrid modes lie within the measurement bandwidth, their superposition can again lead to a Phantom-Drude response (Fig.~\ref{fig:fig4}d). Fitting such a broadened spectrum with a Drude model yields apparent parameters $\gamma_{\text{Phantom}} = \SI{0.42}{THz}$ and $D_{\text{Phantom}} = \SI{63.44}{mS\times THz \times rad}$ that deviate by over \SI{400}{\%} from the true intrinsic values of $\gamma = \SI{0.1}{THz}$ and $D = \SI{43.35}{mS\times THz \times rad}$ (see SI).

These findings highlight that, while low-conductive gates may initially seem like a straightforward choice for on-chip THz spectroscopy, they can complicate data interpretation by strongly modifying the response of a vdW material and potentially giving rise to a Phantom-Drude response. Such gates are most effective when intentionally used for cavity control protocols, provided they are carefully engineered in thickness and dimensions to limit the number of hybridizing self-cavity modes within the experimental sensitivity range \cite{kipp2024cavity}.

In contrast, to probe the intrinsic response of a gate-tunable vdW material with minimal disturbance, a highly conductive gate, such as a \SIrange{20}{30}{nm} thick graphite flake, separated from the 2D material by a $\sim$~\SI{100}{nm} hBN spacer and embedded in a transmission line of design (b) provides an effective solution. In this ``sensing cavity" configuration \cite{kipp2024cavity}, coupling and linewidth broadening are strongly suppressed, and the gate contribution can be reliably subtracted using the on-chip THz architecture with built-in referencing (Fig.~\ref{fig:fig1}a) \cite{Katy2025}. Graphite flakes are widely employed in high-quality DC transport experiments for their reliability and low disorder (Fig.~S11 of Ref.~\cite{zibrov2017tunable}) and allow symmetric, precise carrier density control for both electron and hole doping down to millikelvin temperatures. This is advantageous, as it indicates that the most effective approach for on-chip THz spectroscopy is to use standard graphite gates for vdW heterostructures, while carefully considering boundary conditions and flake thicknesses. \\


\paragraph{\textbf{Discussion and outlook.}}
\label{sec:disc}

By demonstrating that Phantom-Drude responses can emerge in on-chip THz spectra due to choices in sample size, transmission line geometry, vdW material, gate shapes, and gate material, this article highlights an important response regime. We provide design guidelines to enable selective probing of either Drude behavior or cavity electrodynamics in gate-tunable vdW heterostructures, both of which are well suited for extracting intrinsic material parameters. Although this article focused on responses of metallic 2D materials, the same principles extend to systems supporting other collective modes such as phonons, excitons, or plasmons in strange metals and superconductors \cite{marioskipp2025}.

Looking ahead, the analytical framework developed for 2D materials in on-chip THz spectroscopy \cite{marioskipp2025} opens new avenues for tailoring sample designs to probe intricate spectral signatures of emergent phenomena in 2D quantum materials. Importantly, while our discussion centers on a specific on-chip THz spectroscopy circuit featuring two parallel metal traces, the underlying design principles for vdW heterostructures are broadly applicable. They can be readily adapted to other on-chip THz circuit architectures and may provide valuable guidance for a wide range of near-field techniques.

\subsection*{Funding}

M.H.M., H.M.B. and M.W.D. acknowledge support from the Alexander von Humboldt Foundation. H.M.B. acknowledges financial support from the European Union under the Marie Sklodowska-Curie Grant Agreement no. 101062921 (Twist-TOC). G.K. acknowledges support by the German Research Foundation through the Cluster of Excellence CUI: Advanced Imaging of Matter (EXC 2056, project ID 390715994). We acknowledge support by the Deutsche Forschungsgemeinschaft (DFG, German Research Foundation) - 508440990 and  531215165 (Research Unit ‘OPTIMAL’). We acknowledge support from the Max Planck-New York Center for Non-Equilibrium Quantum Phenomena. This research was developed with funding from the Defense Advanced Research Projects Agency (DARPA) under the QUAMELEON Advanced Research Concept. The views, opinions and/or findings expressed are those of the authors and should not be interpreted as representing the official views or policies of the Department of Defense or the U.S. Government.

\subsection*{Acknowledgements}
We acknowledge fruitful discussions with Dante Kennes.

\vspace*{-2mm}
\subsection*{Competing interests} 
The authors declare no competing interests.

\subsection*{Data, materials and code availability}
Data available upon request.

\subsection*{Author contributions}

G.K. and J.W.M. conceived the idea of the project. G.K. performed the analytical theory calculations using the theory developed by M.H.M. with support of G.K., A.M.P. and T.M.. T.M. performed the full 3D electromagnetic simulations with the aid of G.M.. H.M.B, M.H.M., M.W.D and A.M.P participated in the evolution of the ideas and impact within the field. G.K. visualized the data with support from M.W.D.. G.K. wrote the initial manuscript with support in reviewing and editing by all authors. J.W.M. supervised the overall project with support in mentoring from H.M.B.. 

\newpage
\onecolumngrid

\section{Supplement Note 1: Analytical theory}

\subsection{General principle}

Here, we summarize the key ideas of the analytical theory used in the main text. A detailed discussion may be found in Refs.~\cite{kipp2024cavity,marioskipp2025}.

\begin{figure*}[b!]
    \centering
    \includegraphics[width=1. \linewidth]{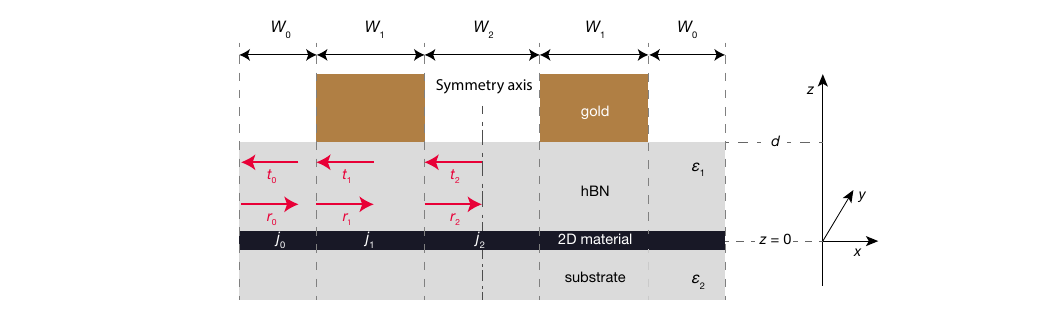}    \caption{\textbf{Cross-section of an on-chip THz device:} A 2D material encapsulated with an insulating layer such as hBN is placed on an insulating substrate. The metal traces that form the transmission lines are deposited on top. The analytical theory describes the effective near-field conductivity and current density profiles by solving Maxwell's equations for transmitted $t_{0,1,2}$ and reflected $r_{0,1,2}$ waves at the dielectric boundaries provided by the sample's edges and metal trace edges (grey dashed lines).}
    \label{fig:samplegeometry}
\end{figure*}

The theory solves Maxwell's equations to describe the electrodynamics of 2D materials embedded in on-chip THz spectroscopy with a cross-section as shown in Fig.~\ref{fig:samplegeometry}. A symmetric sample can be divided into five regions, where $W_0$ is the width of the two unscreened 2D material regions extending beyond the transmission line, $W_1$ is the width of each metal trace and $W_2$ is the width of the transmission line gap. Plasmons are driven by the THz pulse’s electric field throughout all regions of the 2D material, with different dispersion relations for screened and unscreened plasmons (see main text). By solving Maxwell’s equations for the transmitted ($t_{0,1,2}$) and reflected ($r_{0,1,2}$) waves at the dielectric boundaries indicated by the grey dashed lines in Fig.~\ref{fig:samplegeometry}, and assuming continuity of the current density ($j_{0,1,2}$) and electrostatic potential across these boundaries, as well as vanishing current density ($j_0=0$) at the sample edges, one can calculate both the effective plasmonic current-density profile in the integrated 2D material and its corresponding near-field optical conductivity.

Depending on sample size, transmission line design and the conductivity of the sample, the effective near-field optical conductivity approaches a limit, where it resembles a Drude response, or it is characterized by multiple finite-momentum plasmonic resonances, as quantified by the feedback factor $F$. When $F \approx 0$, a Drude response is observed. When $F \neq 0$, the system exhibits a plasmonic cavity or Phantom-Drude response. The expression for $F$ is given in the main text for $W_0 = 0$. For $W_0 \neq 0$, it is given by:

\begin{equation}
    F=\frac{\text{sinc}\left(q_{\text{un}}W_2/2 \right)}{-\cos(\frac{q_{\text{un}}W_2}{2})+\frac{q_{\text{sc}}(q_{\text{un}}\cos(q_{\text{sc}}W_1)\sin(q_{\text{un}}W_0)+q_{\text{sc}}\cos(q_{\text{un}}W_0)\sin(q_{\text{sc}}W_1))\sin(q_{\text{un}}W_2/2)}{q_{\text{un}}(q_{\text{sc}}\cos(q_{\text{un}}W_0)\cos(q_{\text{sc}}W_1)-q_{\text{un}}\sin(q_{\text{un}}W_0)\sin(q_{\text{sc}}W_1))}}.
\end{equation}

\subsection{Current density profiles of inset of Fig.~1b}

\begin{figure*}
    \centering
    \includegraphics[width=1. \linewidth]{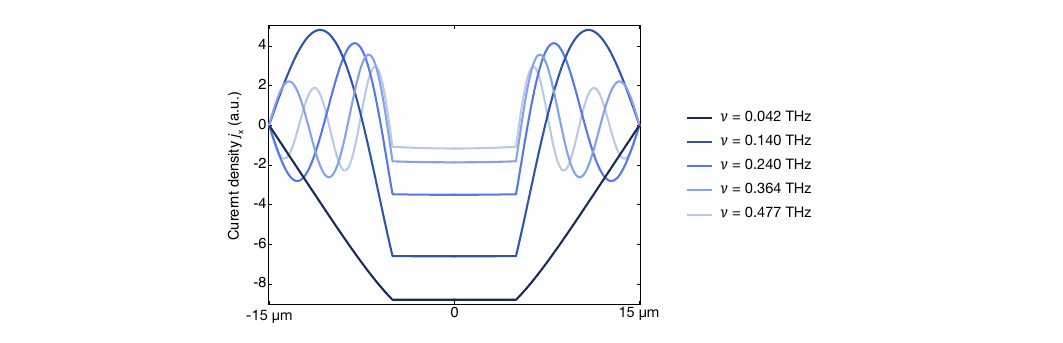}    \caption{\textbf{Current density profiles of the plasmonic resonances shown in Fig.~1b:} The calculations of the current density profiles are calculated with the analytical theory for the five lowest-frequency resonances.}
    \label{fig:currdists_1b}
\end{figure*}

In Fig.~\ref{fig:currdists_1b}, the analytical theory is used to calculate the current density profiles of the plasmonic resonances underlying Fig.~1 of the main text, whose superposition gives rise to the Phantom-Drude response.

\subsection{Symmetric and antisymmetric modes in gated heterostructures}

In case of a heterostructure comprising a 2D material and an electrostatic gate, plasmons are excited in both the 2D material and the gate. Due to the close proximity of the two layers, the plasmons hybridize strongly to form symmetric (in-phase) and antisymmetric (out-of-phase) modes. As a consequence, symmetric and antisymmetric modes form in each region, with different dispersion relations for screened and unscreened modes. For a detailed derivation of the analytical framework describing this dual-layer structure, see Refs.~\cite{kipp2024cavity,marioskipp2025}.

The calculations in Fig.~4b of the main text were performed using the following formulas:

\begin{align}
    \nu_{sc,sym} =& \sqrt{\frac{2 q^2d_1 d_2 D_T D_B}{(2\pi)^2  \left( (d_2 \epsilon_0\epsilon_1 D_B + 
    d_1 \epsilon_0\epsilon_1 (D_T + 
       D_B)) + \sqrt{(4 d_1 d_2 \epsilon_0^2\epsilon_1^2 D_T D_B - 
(d_2 \epsilon_0\epsilon_1 D_B + d_1 \epsilon_0\epsilon_1 (D_T + D_B))^2) } \right)}} ,\label{eq:qsc1} \\
    \nu_{sc,antisym} =& \sqrt{\frac{2 q^2d_1 d_2 D_T D_B}{(2\pi)^2  \left( (d_2 \epsilon_0\epsilon_1 D_B + 
    d_1 \epsilon_0\epsilon_1 (D_T + 
       D_B)) - \sqrt{(4 d_1 d_2 \epsilon_0^2\epsilon_1^2 D_T D_B - 
(d_2 \epsilon_0\epsilon_1 D_B + d_1 \epsilon_0\epsilon_1 (D_T + D_B))^2) } \right)}},\label{eq:qsc2}\\
\nu_{vdW} =& \sqrt{\frac{q^2D_{T}}{(2\pi)^2 \epsilon_0\epsilon_1\left( \frac{1}{d_{1}}+\frac{1}{d_{2}}\right)}} ,\label{eq:qsc3}\\
\nu_{gate} =& \sqrt{\frac{q^2D_{B}(d_{1}+d_{2})}{(2\pi)^2 \epsilon_0\epsilon_1}} ,\label{eq:qsc4}
\end{align}

\noindent
where $\nu_{sc,sym}$ is the resonance frequency of the screened symmetric mode, $\nu_{sc,antisym}$ is the resonance frequency of the screened antisymmetric mode, $\nu_{vdW}$ is the resonance frequency of the uncoupled vdW material plasmon, screened from the metal traces above and from the gate below, $\nu_{gate}$ is the resonance frequency of the uncoupled gate plasmon, screened from the metal traces above, with the momentum $q=2\pi/\lambda$, the top hBN thickness $d_1$, the bottom hBN thickness $d_2$, the Drude weight of the top layer (vdW material) $D_T=D_{\text{vdW material}}$, the Drude weight of the bottom layer (gate) $D_B=D_{\text{gate}}$, the dielectric constant of hBN $\epsilon_1$ and the vacuum permittivity $\epsilon_0$. The following values were used for the calculation in Fig.~4b of the main text: $q=\pi/(\SI{39}{\micro\meter})$, $d_1=\SI{20}{nm}$, $d_2=\SI{100}{nm}$, $D_T=$~\SI{43.35}{mS\times THz \times rad}, $\epsilon_1=3.7$.

The calculations in Fig.~4c,d of the main text were performed using the full analytical theory of dual-layer structures discussed in Refs.~\cite{kipp2024cavity,marioskipp2025}.

\subsection{Ohmic contact within the analytical theory framework}

The ohmic contact scenario can be captured by allowing charge flow between the 2D material and the metal traces in the $W_1$ regions. This is achieved by imbuing the layer separating the 2D material and gold described by the permittivity $\epsilon_1$ with a conductive response. For extremely thin hBN layers for example, the conductive response can represent tunneling between the material and the metal along the $z$-direction. In the $W_1$ regions, the permittivity $\epsilon_1$ is then replaced by:

\begin{equation}
    \epsilon_{1,\text{ohmic}} = \epsilon_1 + \frac{\sigma_{\text{DC,z}}}{-i\omega}. 
\end{equation}

At very small frequencies or equivalently very large $\sigma_{\text{DC,z}}$, such that $\frac{\sigma_{\text{DC,z}}}{- i \omega} \rightarrow \infty $, we reach the conductivity limit, where $q_{sc}$ obtains a large imaginary part, causing plasma waves traveling in the screened region to quickly decay due to the ohmic contact. Mathematically, this causes $\cos(q_2 W_2/2) \propto e^{- \text{Im} \{ q_{2} \} W_2/2 } \rightarrow 0 $. In this case, the feedback factor, for $W_2 = 0 $, $|q_{un}| \ll |q_{sc}|, 2/W_2$ is found to be:

\begin{equation}
    F_{\text{ohmic}} \stackrel{\sigma_{\text{DC,z}} \rightarrow \infty }{\approx} \frac{2}{W_2 q_1} \propto \frac{1}{\sqrt{\sigma_{\text{DC,z}}}},
\end{equation}

\noindent
which vanishes as the inverse $\sqrt{\sigma_{\text{DC,z}}}$. In the perfect ohmic contact case, one can imagine that the 2D material is not separated from the metal traces and is instead directly contacted to the gold. The large conductivity of the gold suppresses $F$ and leads to $\sigma_{\text{near-field}} \approx \sigma_{\text{Drude}}$.

\subsection{Simulation of an irregularly shaped sample within the analytical theory framework}

While the irregularly shaped sample discussed in Fig.~3 of the main text was analyzed through full 3D electromagnetic simulations, the analytical theory can also be used to approximate the effective conductivity of an irregularly shaped sample:

\begin{equation}
    \sigma_{\text{irregularly shaped sample}}=\int \sigma_{\text{near-field}}(\nu,W_0)P(W_0)dW_0,
\end{equation}

\noindent
where $\sigma(\nu,W_0)$ is the frequency- and $W_0$-dependent near-field conductivity and $P(W_0)$ is the distribution of the width $W_0$ across the device.

\section{Supplement Note 2: Full 3D electromagnetic simulations}
\subsection{Method}

The full 3D electromagnetic simulations are carried out with the frequency-domain solver in CST Microwave Studio\cite{CST}, which employs the finite-element method (FEM) to solve Maxwell’s equations. For more details about the method, see supplementary information of Refs. \cite{kipp2024cavity,marioskipp2025}.

Importantly, the simulations account for the effective near-field conductivity of the integrated heterostructure and how efficiently a THz pulse interacts with the sample, thereby closely mimicking experimental conditions. This involves the consideration of a ``filling factor", as discussed in the following section.

\subsection{Filling factor}

While a more detailed theoretical explanation of the filling factor is given in Refs.~\cite{kipp2024cavity,marioskipp2025}, we provide here a purely intuitive picture. The filling factor description is based on the fact that the detected near-field conductivity amplitudes from transmission data depends on how effectively the THz pulse interacts with a 2D material. The degree of interaction, governed by both the transmission line geometry and the thickness of the insulating layer between the transmission line and the 2D material, determines how much of the THz field is modified by the sample. A material placed in direct proximity to the transmission line significantly attenuates the transmitted THz pulse, whereas the same material separated by, for example, a thick hBN layer induces a markedly smaller transmission change. This filling factor thus emerges as a fundamental parameter: without accounting for it, neither Drude conductivities nor spectral weights of plasmonic resonances can be quantitatively inferred from experimental data.

We define the filling factor $C_{\text{filling}}$ as (see Ref.~\cite{kipp2024cavity})

\begin{equation}
    C_{\text{filling}}=\frac{\sigma_{\text{sim}}}{l\sigma_{\text{near-field}}},
\end{equation}

\noindent
with the simulated conductivity $\sigma_{\text{sim}}$, the sample length $l$ in $y$-direction and the near-field conductivity (as calculated by the analytical theory) $\sigma_{\text{near-field}}$.

\subsection{Simulations of ohmic contact design}

As the analytical theory describes samples with an insulating layer between sample and transmission line, full 3D electromagnetic simulations were performed for a transmission line design in which the sample is placed in direct ohmic contact (see Fig.~2, main text). The simulations considered a sample with Drude weight $D = \SI{6.28}{mS\times THz\times rad}$ and scattering rate $\gamma = \SI{0.1}{THz}$, for various sample widths $W=2W_1+W_2$. 

\begin{figure*}
    \centering
    \includegraphics[width=1. \linewidth]{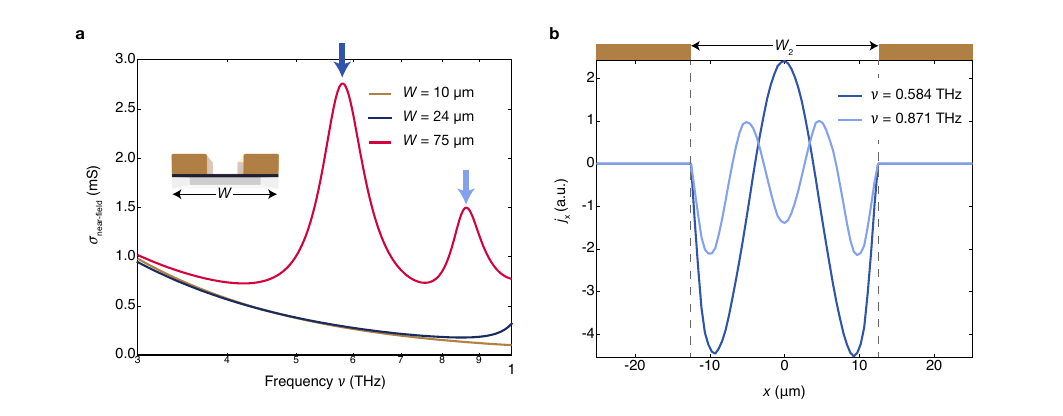}    \caption{\textbf{Current density profiles of the plasmonic resonances shown in Fig.~2d for the ohmic contact design: a} Full 3D electromagnetic simulations of the near-field optical conductivity as function of frequency for three different sample widths. Whereas Fig.~2d of the main text displays the response over the full \SIrange{0.001}{1}{THz} range, the plot shown here focuses on the resonances near the upper end of the bandwidth. Note that to extract $\sigma_{\text{near-field}}$ from the simulations, the following filling factors were considered: $C_{\text{filling}}(W=\SI{10}{\micro\meter})=\SI{326}{mm^{-1}}$, $C_{\text{filling}}(W=\SI{24}{\micro\meter})=\SI{139}{mm^{-1}}$, $C_{\text{filling}}(W=\SI{75}{\micro\meter})=\SI{41}{mm^{-1}}$. \textbf{b} Simulated current density profiles as a function of $x$ for the two resonances highlighted with the arrows in panel (a). The low-frequency resonance stems from an unscreened plasmon with wavelength $\lambda\approx \frac{2}{3}W_2$, while the higher frequency resonance arises from an unscreened plasmon that forms in the transmission line gap with $\lambda\approx \frac{2}{5}W_2$.}
    \label{fig:ohmiccontact_currdists}
\end{figure*}

As shown in Fig.~\ref{fig:ohmiccontact_currdists}, a Drude response is observed for all sample widths. In addition, plasmonic resonances emerge on top of the Drude background for larger sample widths. These correspond to unscreened plasmons that form within the transmission line gap. As shown in Fig.~\ref{fig:ohmiccontact_currdists}b, the two lowest frequency resonances correspond approximately to $\lambda\approx \frac{2}{3}W_2$ and $\lambda\approx \frac{2}{5}W_2$ plasmons. The resonance frequencies of the unscreened plasmonic modes can therefore be approximated with:

\begin{equation}
    \nu=\sqrt{\frac{D}{2\pi\lambda\epsilon_0(\epsilon_1+\epsilon_2)}}.
\end{equation}

\noindent
with wavelengths $\lambda\approx 2W_2/a$, with $a=3,5,\dots$. However, this approximation neglects the boundary conditions imposed by the ohmic contact to the transmission-line metal, resulting in deviations on the order of \SI{100}{GHz} in the predicted resonance frequencies.

As the resonance frequencies of unscreened plasmons are much higher than those of screened or hybrid screened/ unscreened modes, they enter the experimental bandwidth only for sufficiently large sample widths or for samples with low conductivity. This has important implications for sample design. Depending on the goal, one can choose the geometry for a given conductivity such that the plasmonic resonances lie outside the measurement bandwidth, ensuring a purely Drude response, or intentionally bring one plasmonic mode into the accessible range. In the latter case, the presence of a resolvable plasmonic resonance provides an additional tool for extracting intrinsic material parameters. This is particularly beneficial for samples with low scattering rates, where a significant fraction of the spectral weight lies below the measurement bandwidth and would otherwise be difficult to quantify. In contrast, if the sample parameters are not considered prior to fabrication, there is a risk of obtaining spectra with multiple overlapping plasmonic resonances in addition to the Drude response, resulting in a broad Phantom-Drude response that cannot be used to reliably extract intrinsic material properties.

\subsection{Simulations of an irregularly shaped sample}

\begin{figure*}
    \centering
    \includegraphics[width=1. \linewidth]{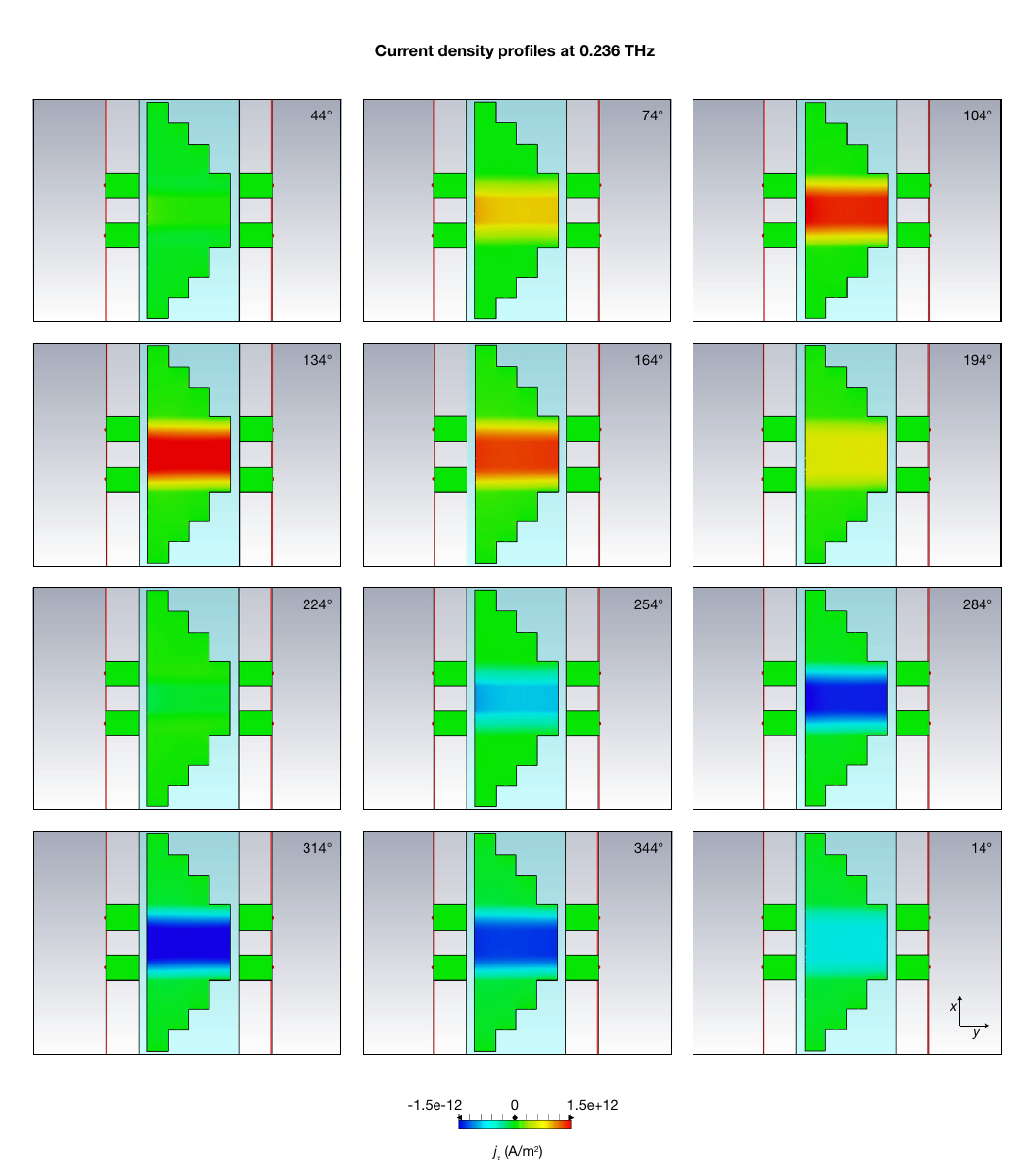}    \caption{\textbf{Current density distributions of the irregularly shaped device analyzed in Fig.~3 at \SI{0.236}{THz}, shown at different phases:} Full 3D electromagnetic simulations of the current density along the $x$-direction, perpendicular to the THz pulse propagation, at the frequency of the lowest-frequency plasmonic mode. In this mode, the current density in the $W_0$ regions remains near zero at all phases, resulting in minimal inhomogeneous broadening of the lowest-frequency resonance of the irregularly shaped sample.}
    \label{fig:currdists_irr_236GHz}
\end{figure*}

\begin{figure*}
    \centering
    \includegraphics[width=1. \linewidth]{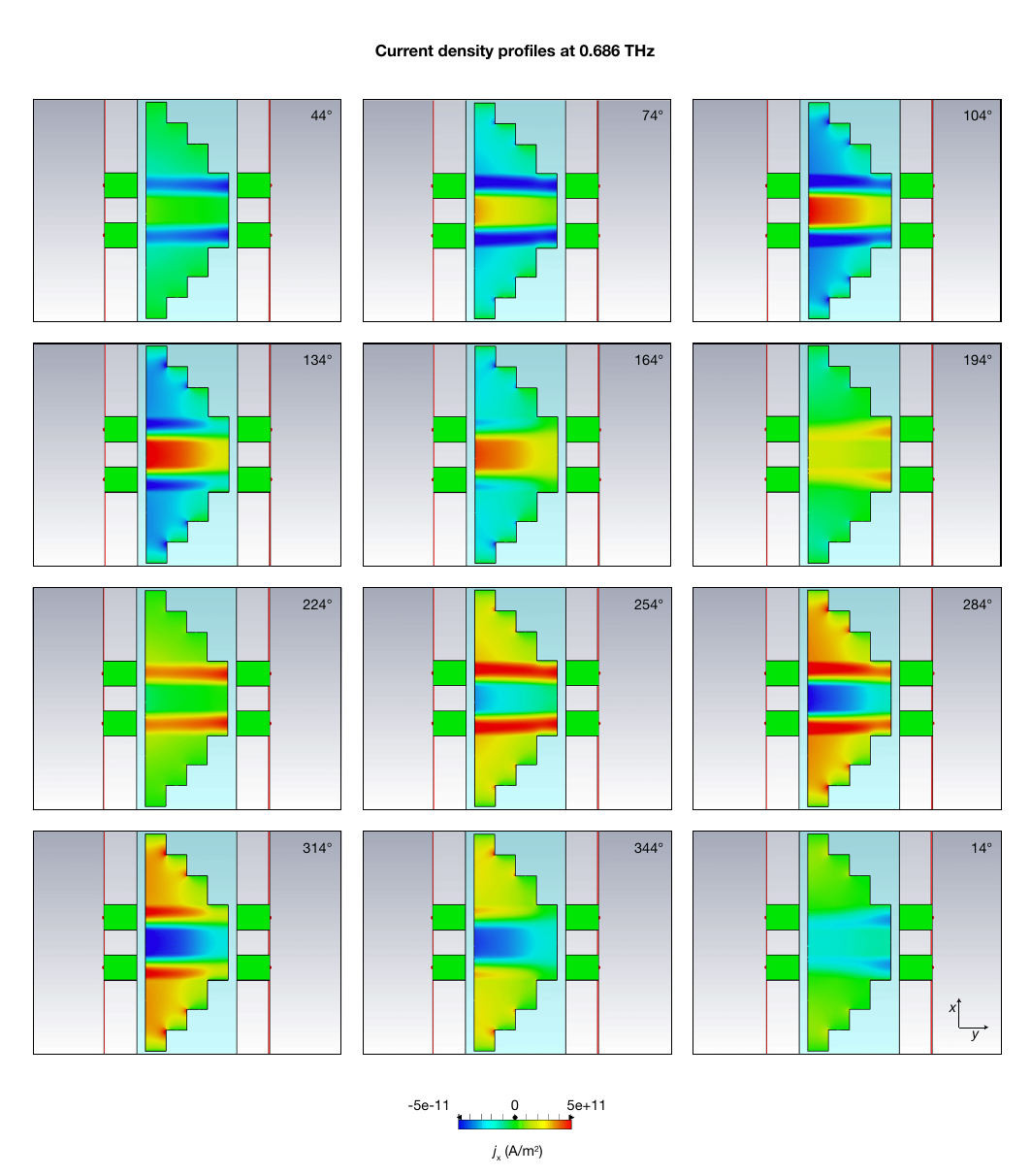}    \caption{\textbf{Current density distributions of the irregularly shaped device analyzed in Fig.~3 at \SI{0.686}{THz}, shown at different phases:} Full 3D electromagnetic simulations of the current density along the $x$-direction, perpendicular to the THz pulse propagation, at the frequency of the second-lowest-frequency plasmonic mode. In this mode, the current density in the $W_0$ regions becomes comparable to that in the $W_1$ and $W_2$ regions. Since the unscreened plasmons in the $W_0$ regions have different wavelengths, and therefore resonance frequencies that depend on $W_0$, the second-lowest-frequency resonance of the irregularly shaped sample exhibits pronounced inhomogeneous broadening (see Fig.~3).}
    \label{fig:currdists_irr_686GHz}
\end{figure*}

To accurately capture the response of the irregularly shaped sample, full 3D electromagnetic simulations were performed for the structure shown in Fig.~2 of the main text, in addition to the individual slices of the structure. A sample length of $l=\SI{10}{\micro\meter}$ (in $x$-direction) was chosen so that each section with a different total width has a length of \SI{2.5}{\micro\meter}. The current-density profiles of the $x$-directed current $j_x$ at the two resonance-peak frequencies are shown for different phases in Figs.~\ref{fig:currdists_irr_236GHz} and \ref{fig:currdists_irr_686GHz}.

As shown in Fig.~\ref{fig:currdists_irr_236GHz}, the current density in the $W_0$ regions extending beyond the transmission line is essentially zero. Consequently, the resonance frequency of the lowest-momentum plasmonic mode is largely insensitive to the width $W_0$, and this mode does not experience significant inhomogeneous broadening due to the sample’s irregular shape (see low-frequency mode in Fig.~3 of the main text).

In contrast, the higher-momentum modes exhibit a nonvanishing current density in the $W_0$ regions. As illustrated in Fig.~\ref{fig:currdists_irr_686GHz}, the plasmonic excitation spans the full sample width in the $x$-direction, with comparable current amplitudes in the $W_0$, $W_1$, and $W_2$ regions. As a result, these higher-momentum modes are strongly affected by the sample geometry, leading to pronounced inhomogeneous broadening in non-rectangular samples (see high-frequency mode in Fig.~3 of the main text).

\subsection{Simulations of low-conductive samples}
\begin{figure*}
    \centering
    \includegraphics[width=1. \linewidth]{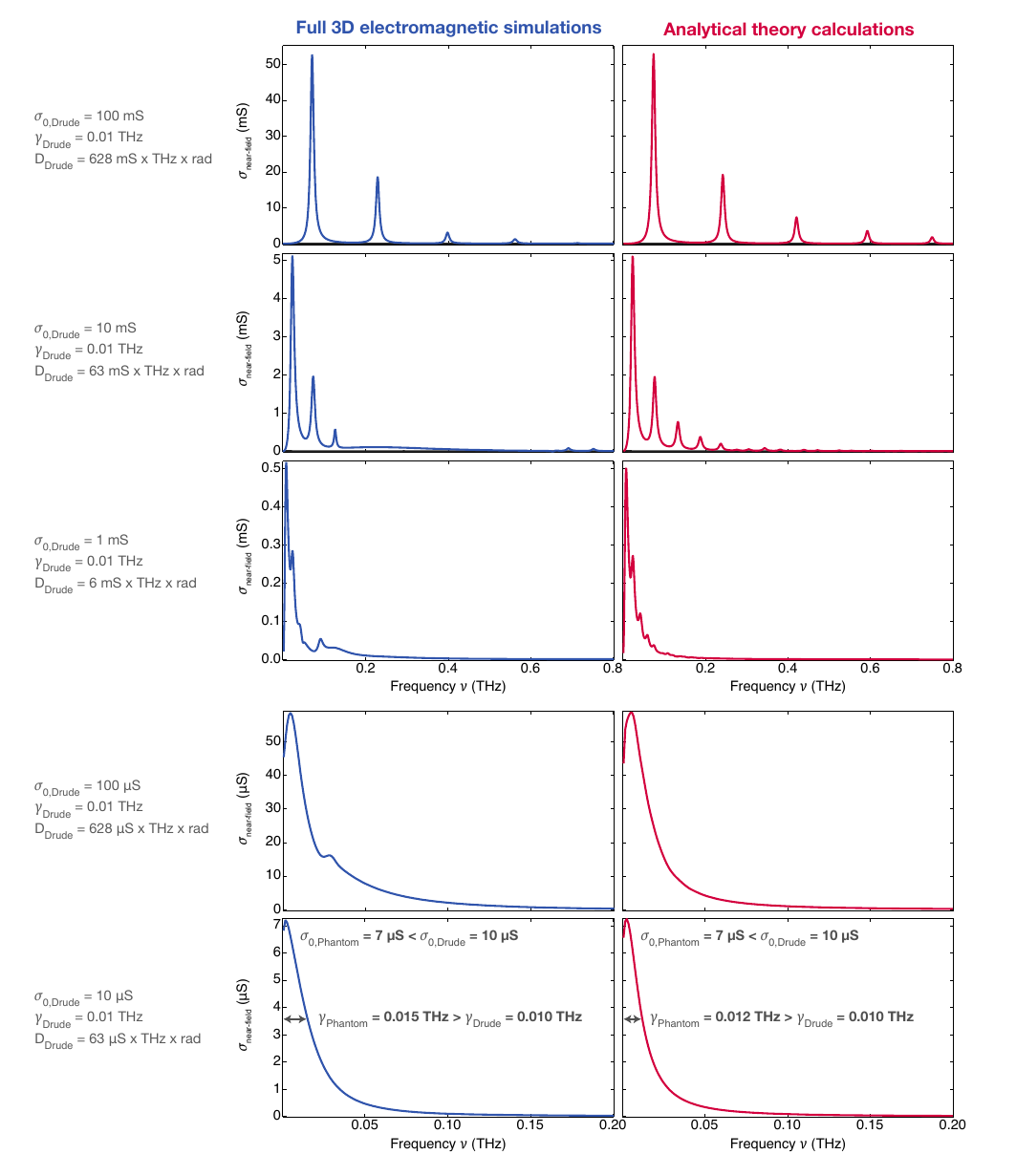}    \caption{\textbf{Full 3D electromagnetic simulations and analytical theory calculations of a sample with varying Drude weight:} A conductive layer with varying Drude weight $D_{\text{Drude}}=\sigma_0\times \gamma$ and a fixed scattering rate  $\gamma=\SI{0.01}{THz}$ is embedded in a heterostructure with $d_{\text{hBN}}=\SI{100}{nm}$, $W_0=0$ and $W_1=W_2=\SI{10}{\micro\meter}$. The system is analyzed using both full 3D electromagnetic simulations and analytical theory calculations. At the lowest simulated Drude weight, $D_{\text{Drude}}=\SI{63}{\micro S \times THz \times rad}$, the plasmonic resonances are shifted to sufficiently low frequencies that they overlap, resulting in a superposition that manifests as a Phantom-Drude response.}
    \label{fig:SI_lowconductivesamples}
\end{figure*}

To validate the analytical theory and the conditions under which a phantom-Drude regime can emerge across a wide range of conductivities (or Drude weights), full 3D electromagnetic simulations were performed for a structure with $d_{\text{hBN}}=\SI{100}{nm}$, $W_0=0$ and $W_1=W_2=\SI{10}{\micro\meter}$, while systematically varying $D_{\text{Drude}}$. As shown in Fig.~\ref{fig:SI_lowconductivesamples}, the analytical theory and full 3D electromagnetic simulations exhibit good agreement over a broad range of Drude weights. Notably, plasmonic resonances persist even at low Drude weights, such as $D_{\text{Drude}}=\SI{63}{\micro S \times THz \times rad}$. As shown in the bottom panels of the figure, these resonances overlap in frequency, giving rise to a phantom-Drude response for the chosen sample geometry.

\newpage\clearpage
\bibliography{references}

\end{document}